\documentclass[aps,prl,revtex4-1,reprint,twocolumn,superscriptaddress,showpacs,letter]{revtex4-1}
\usepackage{hyperref}
\usepackage{natbib}
\usepackage{amsmath}
\usepackage[]{graphicx}
\newcommand{\bra}[1]{\left\langle {#1} \right|}
\newcommand{\ket}[1]{\left|  #1 \right\rangle}
\newcommand{\aver}[1]{\langle {#1} \rangle}

\newcommand{\raw}[0]{\rightarrow}

\begin{document}

\title{Suppression of Ion Transport due to Long-Lived Sub-Wavelength Localization by an Optical Lattice}

\author{Leon Karpa}
\thanks{These authors have contributed equally to this work.}
\affiliation{Department of Physics and Research
Laboratory of Electronics, Massachusetts Institute of Technology, Cambridge,
Massachusetts 02139, USA}

\author{Alexei Bylinskii}
\thanks{These authors have contributed equally to this work.}
\affiliation{Department of Physics and Research Laboratory of Electronics, Massachusetts Institute of Technology, Cambridge, Massachusetts 02139, USA}

\author{Dorian Gangloff}
\thanks{These authors have contributed equally to this work.}
\affiliation{Department of Physics and Research Laboratory of Electronics, Massachusetts Institute of Technology, Cambridge, Massachusetts 02139, USA}
\thanks{These authors contributed equally to this work.}

\author{Marko Cetina}
\affiliation{Department of Physics and Research Laboratory of Electronics, Massachusetts Institute of Technology, Cambridge, Massachusetts 02139, USA}
\affiliation{Institute of Quantum Optics and Quantum Information, Otto-Hittmair-Platz 1, A-6020 Innsbruck, Austria}

\author{Vladan Vuleti\'c}
\email[]{vuletic@mit.edu}
\affiliation{Department of Physics and Research Laboratory of Electronics, Massachusetts Institute of Technology, Cambridge, Massachusetts 02139, USA}

\date{\today}

\begin{abstract}
We report the localization of an ion by a one-dimensional optical lattice in the presence of an applied external force. The ion is confined radially by a radiofrequency trap and axially by a combined electrostatic and optical-lattice potential. The ion is cooled using a resolved Raman sideband technique to a mean vibrational number $\aver{n} = 0.6 \pm 0.1$ along the optical lattice. We implement a detection method to monitor the position of the ion subject to a periodic electrical driving force with a resolution down to $\lambda$/40, and demonstrate suppression of the driven ion motion and localization to a single lattice site on time scales of up to 10 milliseconds. This opens new possibilities for studying many-body systems with long-range interactions in periodic potentials.
\end{abstract}
\maketitle
Time-dependent optical forces play an important role in quantum information processing with ions \cite{Wineland13,Leibfried03}, as they provide a means to convert the strong Coulomb interaction into a state-dependent ion-ion interaction \cite{Monroe95,Roos99}. Time-independent periodic optical potentials (optical lattices) have been used to study anomalous ion diffusion \cite{Katori97}, and more recently have been proposed as a way to implement various quantum computation and simulation schemes \cite{Cirac00,Porras04,Porras04a,Schmied08,Friedenauer08, Blatt08a,Haffner08, Kim10, Schneider10,Monz11, Islam11, Blatt12,Schneider12}. An ion chain in an optical lattice represents a many-body system with strong long-range interactions, that is predicted to exhibit both classical and quantum phase transitions in the context of the Frenkel-Kontorova friction model \cite{Pruttivarasin11,Benassi11, Garcia-Mata07}.

Compared to neutral atoms, the optical trapping of ions is substantially more difficult: even weak stray electric fields produce strong Coulomb forces, while in deep optical potentials the associated linewidth broadening thwarts efficient Doppler cooling. Nonetheless, by superimposing a tightly focused optical running-wave dipole trap with a Paul trap and turning the latter off after loading, trapping of a single ion for a few milliseconds has been achieved \cite{Schneider10}. 
In one-dimensional optical lattices, trapping times of tens of microseconds \cite{Enderlein12}, as well as spectroscopic evidence for sub-wavelength localization in the lattice on microsecond time scales, have been reported recently \cite{Linnet12}.

\begin{figure}[h!!!]
  \begin{center}
    \includegraphics [width=9.0cm]{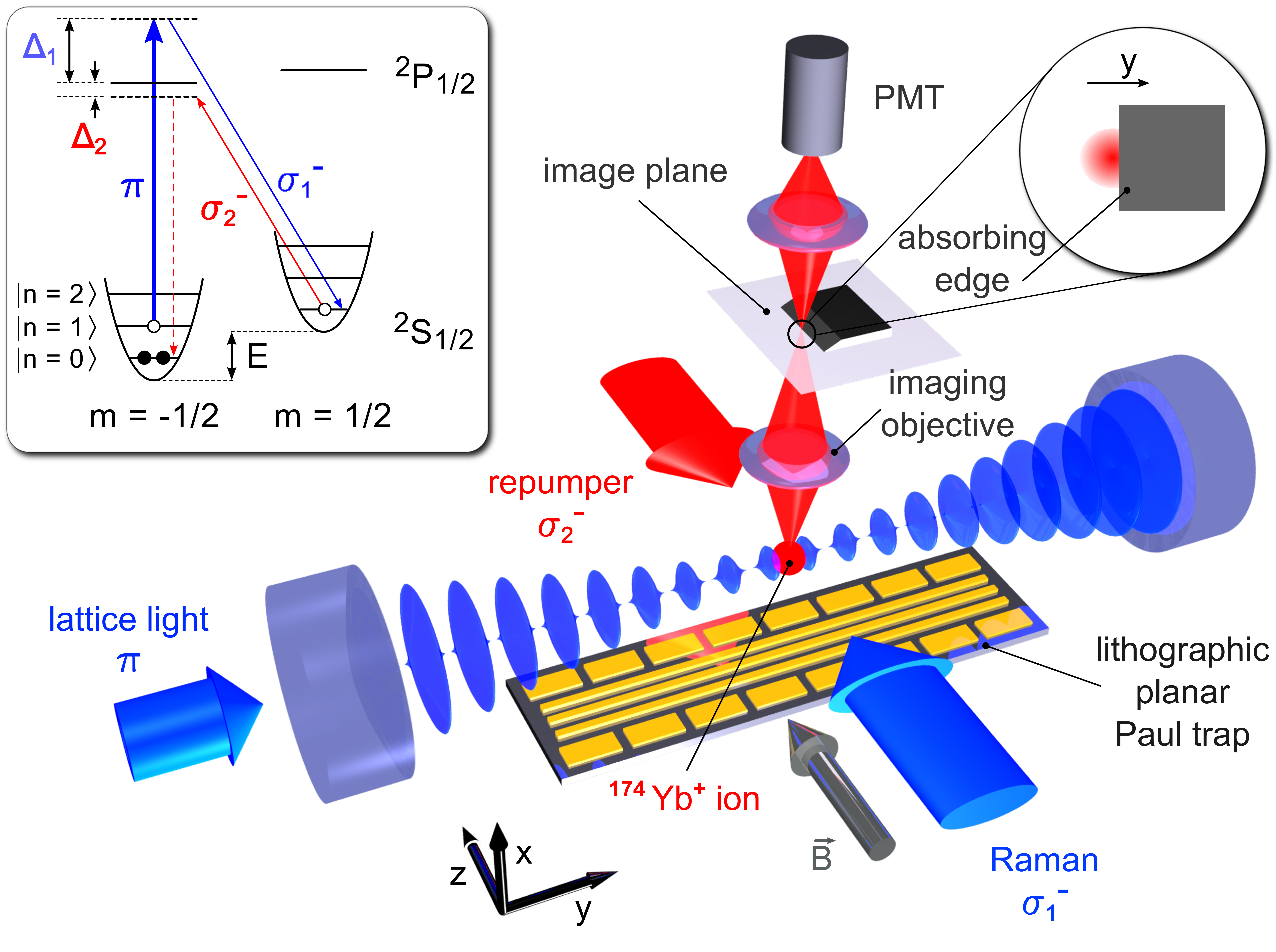}
  \caption{\label{fig:setup} Experimental setup for trapping an ion in an optical lattice over a microchip. An optical resonator is aligned with the linear Paul trap, located 135 $\mu$m above the chip surface. The inset shows the relevant transitions in $^{174}$Yb$^{+}$ ions. A transverse magnetic field $B_{z}$ is applied along the z direction. The far-detuned $\pi$-polarized lattice and Raman $\sigma_{1}^{-}$ beams are used to drive a two-photon transition that reduces the ion's energy, while the near-resonant $\sigma_{2}^{-}$ beam optically pumps the ion into its initial spin state $\ket{^{2}S_{1/2}, m=- \frac{1}{2}}$, completing the cooling cycle. An absorbing edge in the image plane is used to map the ion fluorescence to position, enabling measurements with sub-wavelength resolution.}
  \end{center}
 \end{figure}

In this Letter we demonstrate the long-lived localization of an ion to a single site of an optical lattice even when the ion is subjected to a Coulomb force that in the absence of the optical potential moves the ion by several lattice sites. Measuring the ion's center position with a resolution of 20~nm, well below the 185~nm lattice period, we find that the electric-field-induced ion transport is fully suppressed by the lattice on time scales of up to 10~ms, and suppressed by 30\% at 0.1~s, much longer than the $\sim$1~$\mu$s oscillation period in the optical trap.

The apparatus, whose main features are depicted in Fig.\ref{fig:setup}, is described in more detail in Ref. \cite{Cetina13}. The lithographic planar Paul trap contains electrodes driven at a 16~MHz radiofrequency (rf) (three prolonged central electrodes), providing radial confinement with vibrational frequencies of $\omega_z \approx \omega_x \approx 2\pi \times 1.1$~MHz. Additional dc electrodes are used to shape the electrostatic potential along the $y$-axis. To enhance the laser power for the optical-lattice trap, we make use of an optical cavity with a finesse of $\mathcal{F} \approx 1500$ that is aligned with the linear rf quadrupole trap located at a distance of 135~$\mu$m from the chip surface. The TEM$_{00}$ mode of the cavity has a waist size of 38 $\mu$m, and is tuned into resonance with the lattice laser, which is blue detuned by $\Delta_1 = 2\pi \times 12.7$~GHz from the $^{2}S_{1/2}$ $\longrightarrow$ $^{2}P_{1/2}$ transition in $^{174}$Yb$^{+}$ at a wavelength of $2\pi/k = \lambda = 369$~nm. Both the lattice laser and the optical cavity are actively stabilized to a reference resonator. The typical trap depth is $U/h = 45$~MHz, corresponding to a circulating power of 10~mW inside the cavity, and a trap vibrational frequency of $\omega_y^{(ol)} = 2\pi \times 1.2$~MHz in the optical lattice.

The ion is prepared in a Paul trap with weak electrostatic confinement along the $y$ axis, $\omega_y^{(dc)} = 2\pi \times 130$~kHz, with a superimposed optical lattice. In order to optically cool the ion in the presence of AC Stark shifts of up to $2 U/h$ that exceed the atomic linewidth of $\Gamma = 2 \pi \times 19.6$~MHz of the $^2S_{1/2} \raw$ $^2P_{1/2}$ transition \cite{Olmschenk09}, we employ a lattice-assisted Raman sideband cooling method \cite{Monroe95a} where the trapping light itself is utilized to drive Raman transitions \cite{Vuletic98,Kerman00}. In our cooling scheme, as indicated in the inset of Fig. \ref{fig:setup}, the $\pi$-polarized light of the trap standing wave in combination with a circularly polarized beam $\sigma_1^-$ along the $z$ axis drives a Raman transition between the levels $\ket{^{2}S_{1/2}, m = -\frac{1}{2}}$ and $\ket{^{2}S_{1/2}, m = \frac{1}{2}}$. The beam along $z$ is frequency shifted by $\delta = 2\pi \times 6$~MHz relative to the lattice laser, and an external magnetic field along the $z$ axis is used to tune the Raman transition into resonance with the vibrational sideband $\ket{m = -\frac{1}{2}; n_y,n_z} \rightarrow \ket{m = \frac{1}{2}; n_y - 1,n_z}$, where $n_y$ and $n_z$ indicate the trap vibrational levels in the $y$ and $z$ directions, respectively. In the blue-detuned lattice the atoms are trapped near the nodes where the vibrational coupling by the Raman beams is of the form $\bra{n_y',n_z'}e^{ikz} \sin ky \ket{n_y,n_z}$, so that the only cooling to first order in the Lamb-Dicke parameter $\eta = (\omega_{rec} / \omega_{y}^{(ol)} )^{1/2}$ occurs along $y$, where $\omega_{rec} = 2 \pi \times 8$ kHz is the recoil frequency. On this transition the Rabi frequency is $ \Omega_2\sqrt{n} \eta$, with a typical two-photon Rabi frequency of $\Omega_2 \sim 2 \pi \times 2.5$~MHz (calculated from the measured saturation parameters of the lattice light and the $\sigma_{1}^{-}$ beam).

To complete the cooling cycle, the ion is illuminated with an additional near-resonant $\sigma^{-}$-polarized pumping beam ($\sigma_{2}^{-}$), detuned from the excited state by $\Delta_2 = - 2\pi \times 100$~MHz, that populates the $\ket{m= -\frac{1}{2}}$ state via spontaneous decay. Under ideal conditions, population  accumulates in the dark lowest vibrational level $\ket{m = -\frac{1}{2};n_y=0}$ that is coupled neither by the Raman transition nor by the optical pumping beam.

\begin{figure}[h!]
  \begin{center}
    \includegraphics [width=8cm]{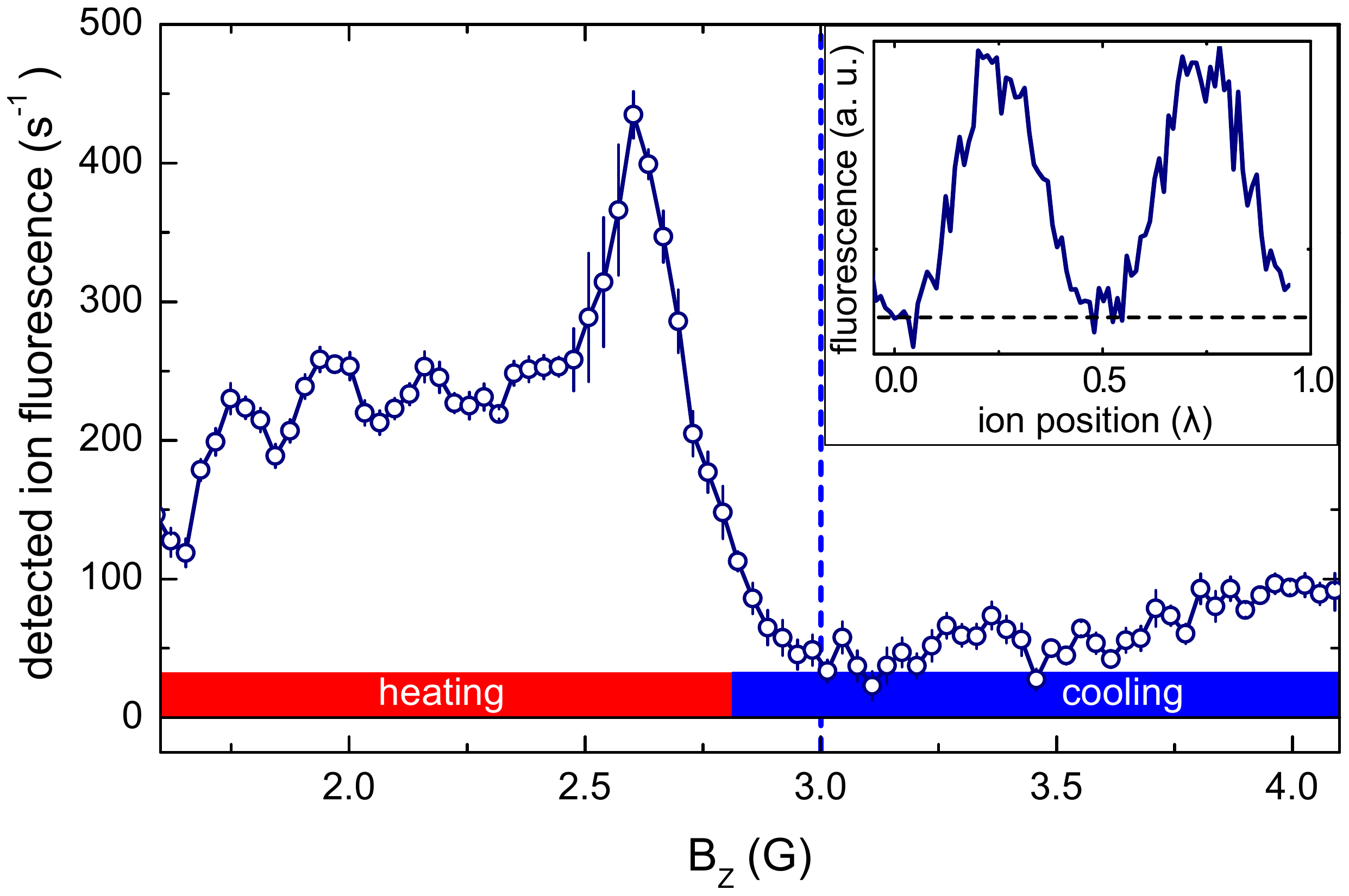}
  \caption{\label{fig:raman} Fluorescence of a Raman cooled ion in an optical lattice with a depth of $U/h = 45$~MHz as a function of an applied magnetic field $B_{z}$ along the $z$ axis which adjusts the two-photon detuning. The background due to light scattering from the chip has been subtracted. The vertical dashed line indicates the predicted location of the red sideband transition at 3 G with an estimated uncertainty of $\pm$0.2 G. The variation in scattering rate originates from the temperature-dependent trap-induced AC Stark shift. The inset shows the florescence from an ion positioned at different locations inside the cavity mode using a strongly confining electrostatic potential, where the dashed line indicates the scattering rate without the optical lattice.}
  \end{center}
 \end{figure}

In the blue-detuned lattice the AC Stark shift at the potential minimum is zero, and increasing temperature along the lattice direction ($y$) results in an AC Stark shift that reduces the detuning of the optical pumping beam ($\sigma_2^-$) from resonance, which manifests itself in a higher scattering rate. (This is illustrated in the inset of Fig. \ref{fig:raman}, where an ion positioned at different locations in the optical lattice by a strongly confining electrostatic potential along $y$ exhibits a spatial variation in fluorescence \cite{Cetina13}.) The scattering rate at the antinodes is significantly increased compared to the nodes. Because of this mechanism, in the optical lattice, the ion's photon scattering rate is a measure of temperature along the lattice direction, and can be used to identify Raman cooling.

Fig. \ref{fig:raman} shows ion fluorescence from a Raman cooled ion in an optical lattice with a trap depth of $U/h = (45 \pm 5)$~MHz. The spectrum is obtained by varying the magnetic field $B_{z}$ along the $z$ axis in order to change the Raman resonance condition. On the heating side of the spectrum, where a Raman transition increases the ion's energy, the ion delocalizes in the optical lattice, which increases the scattering rate by the optical pumping beam. The principal cooling feature of the spectrum is the decrease of fluorescence almost to the zero at approximately $B_{z}$ = 3 G corresponding to the two-photon detuning $\delta_{R} = \mu_{B}B_{z}/ \hbar - ( \delta + \Delta_{ac}/\hbar ) = \omega_y^{(ol)} = 2\pi \times (1.2~ \pm~ 0.1)$~MHz expected for the red sideband of the Raman transition coupling the states $\ket{m = -\frac{1}{2}; n_y}$ and $\ket{m = \frac{1}{2}; n_y -1}$. Here, with $\mu_{B}$ being the Bohr magneton, $\Delta_{ac} / \hbar \approx 2 \pi~ \times (0.7 \pm 0.1)$ MHz at the node, $\mu_{B}B_{z}/ \hbar$ and $\delta = 2\pi \times 6$~MHz denote the differential AC Stark shift between the ground states induced by the optical pumping beam, the Zeeman splitting of these states, and the frequency difference between the two Raman beams, respectively. Since any sizeable extent of the ion's wave function compared to the lattice period of $\lambda/2=185 $~nm would lead to an increase of fluorescence, as observed on the heating side $\delta_R<0$, the fluorescence dip at $\delta_R = \omega_y^{(ol)}$ can be interpreted as a signature of localization in a lattice site to a spatial extent much less than $\lambda/2$.

Under ideal conditions, the fluorescence rate due to Raman transitions is proportional to the fraction of the population that is not hidden in the dark vibrational ground state $\ket{m= - \frac{1}{2}; n_{y} = 0}$, i.e. $ \frac{\aver{n_{y}}}{\aver{n_{y}} + 1}$ \cite{Vuletic98,Hamann98,Perrin98}. Hence, in the regime where $\Omega_{2} \sqrt{\aver{n_{y}} } \eta \gg \Gamma_{op}$, with $\Gamma_{op} = (2.3 \pm 0.3) \times 10^5$ s$^{-1}$ being the optical pumping rate out of the $\ket{m=\frac{1}{2}}$ ground state, the residual observed fluorescence can be used to estimate $\aver{n_{y}}$. The fluorescence rate due to the optical pumping process is given by $3 \frac{\Gamma_{op}} {2}  \frac{\aver{n_{y}}}{\aver{n_{y}} + 1}$, where the factor of 3 arises from the branching ratio of the spontaneous decay from $\ket{2~ P_{1/2}, m = - \frac{1}{2}}$ and the factor 1/2 accounts for two-photon saturation. In this limit, we can determine the mean vibration quantum number as $\aver{n_{y}} = 0.6 \pm 0.1$. Thus the average axial energy of the cooled ion $\aver{E}/h = \left( \frac{1}{2} + \aver{n_{y}} \right) \omega_y^{(ol)} / (2 \pi) = ( 0.7 \pm 0.1 )$~MHz is substantially below the optical-lattice depth of $U/h = 45$~MHz, confirming localization of the ion well below the lattice spacing $\lambda / 2$.

\begin{figure}[]
  \begin{center}
    \includegraphics [width=8cm]{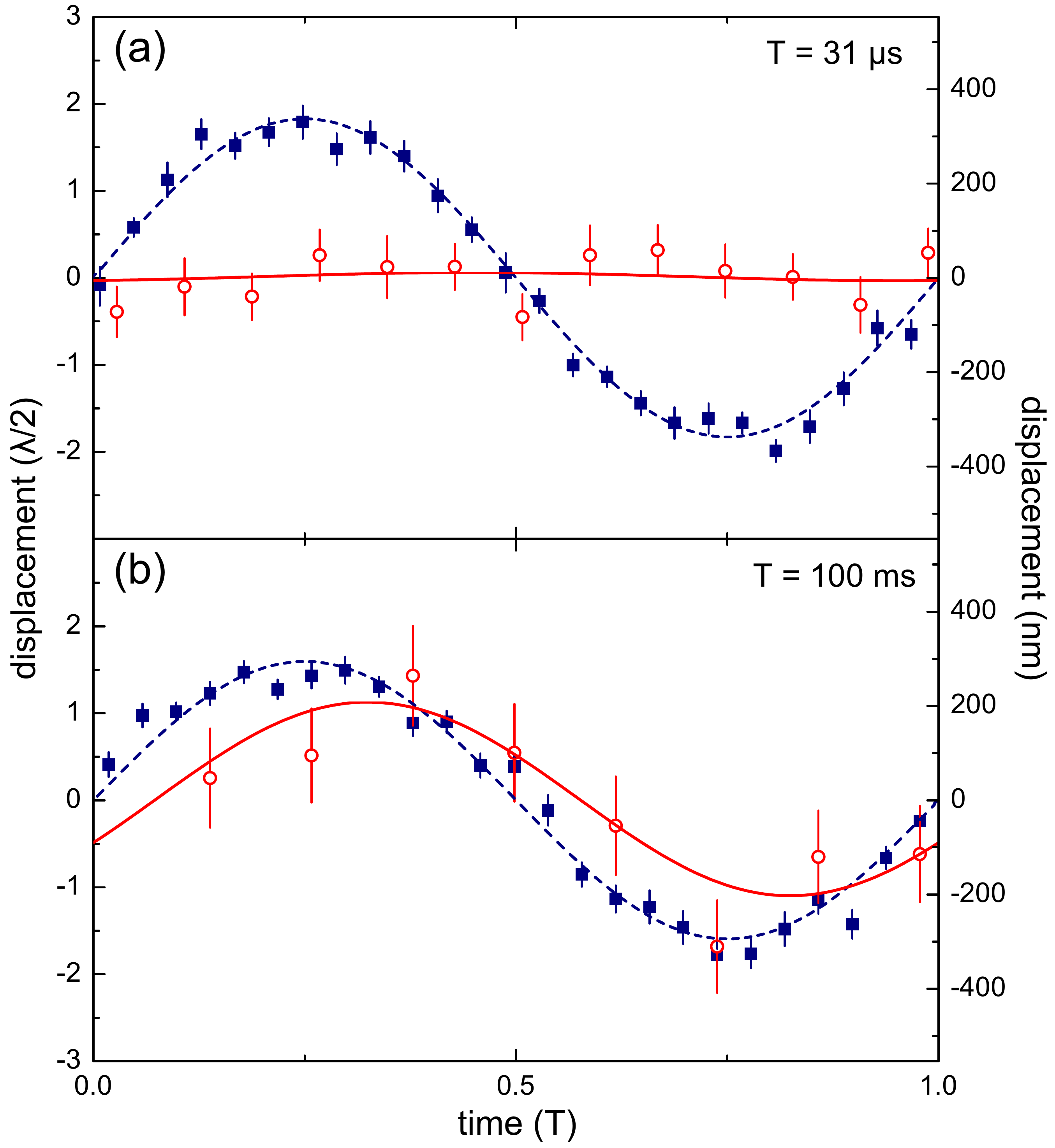}
  \caption{\label{fig:suppression} Ion displacement by a an external electric driving field in units of the lattice site spacing, $\lambda$/2 = 184.5 nm, without optical confinement (squares) and in an optical lattice (dots). (a) The external modulation of the electrostatic potential minimum with a modulation period of T = 31 $\mu$s was applied for averaging durations of 150 s and 300 s respectively. The lines represent sinusoidal fits yielding amplitudes of $A=$ (340 $\pm$ 10) nm in the Paul trap and $A^{(ol)}=$ (10 $\pm$ 20) nm in presence of the optical lattice, respectively. (b) Corresponding measurements for a modulation period of T = 100 ms. The fit yields $A=$ (300 $\pm$ 10) nm, $A^{(ol)}=$ (200 $\pm$ 40) nm and a phase delay $\Delta\phi =$ (25 $\pm$ 11)$^{\circ}$ of the driven motion in the lattice.}
  \end{center}
\end{figure}

Having established the condition for Raman cooling and ion localization in the lattice, we now demonstrate that the optical lattice can indeed suppress ion transport in an externally applied electric field. In order to measure the position of the ion with a resolution below the optical resolution of our imaging system of $D = 2.9~ \mu$m, we position a sharp absorbing edge in the imaging plane such that it blocks half the ion fluorescence as depicted in Fig. \ref{fig:setup}. The remaining fluorescence, as collected on a photomutiplier tube, is then linearly dependent on the ion position, where the measured fractional change in fluorescence of $2.5 \ \times 10^{-4}/$nm is independent of temperature for ions localized to less than $D$, as is the case for all temperatures observed in our system. This fluorescence conversion to a position displacement was performed using an independent measurement with a charge-coupled camera. By sufficiently long signal integration we can find the average position of the ion to much better than $D$ or even the optical wavelength $\lambda$ \cite{Karski09}. Since our photon detection rate at the optimum Raman cooling point is less than 30~s$^{-1}$, and we need many photons for sufficiently high signal-to-noise ratio of the position measurement, we employ a lock-in technique that enables resolving relative ion motion on time scales much shorter than the signal integration time. We also slightly misalign the polarization of the optical pumping beam $\sigma_2^-$ to increase the fluorescence to typically 100~s$^{-1}$. In the Paul trap with low axial frequency $\omega_y^{(dc)} = 2\pi \times 130$~kHz, we apply a slowly varying electric force with period $T$ to displace the ion along the $y$ axis, measure the time-resolved ion fluorescence synchronously with the applied force, and integrate the signal. We choose the amplitude of the driving field to displace the ion by $A \cong \pm ~2$ lattice sites. The corresponding data under Doppler cooling conditions are shown in Fig. \ref{fig:suppression} as the blue solid squares. With an integration time of 150~s, we can resolve the ion's response to the applied force down to 10~nm $\approx \lambda / 40$.

We then turn on the optical lattice and switch to Raman cooling. If the modulation is not too slow, we observe a strong suppression of the ion's driven motion (Fig. \ref{fig:suppression} a), to an amplitude consistent with zero ($A^{(ol)} = (10~ \pm~ 20)$~nm), and much smaller than the lattice spacing $\lambda/2=185$~nm or the amplitude $A=(340 \pm 10)$~nm in the pure Paul trap under the same conditions. For an ion confined to a single site, the expected amplitude of the ion motion is 4~nm in the lattice with vibration frequency $\omega_y^{(ol)} = 2\pi \times 1.2$~MHz, consistent with our observation. Similar results are obtained for driving periods $T$ between $30~ \mu$s and 10 ms. We conclude that on these time scales, much longer than the trap vibrational period of 1~$\mu$s, the ion remains confined to a single lattice site. At still lower driving frequencies ($T \gtrsim 10$~ms) the continuously cooled ion has some finite probability to thermally hop across the lattice potential barrier during the observation time, resulting in a partial suppression of the driven motion when averaged over many trials. This partial amplitude suppression is accompanied by a delay in the ion's motion with respect to the driving force (Fig. \ref{fig:suppression} b).

\begin{figure}[]
  \begin{center}
    \includegraphics [width=9cm]{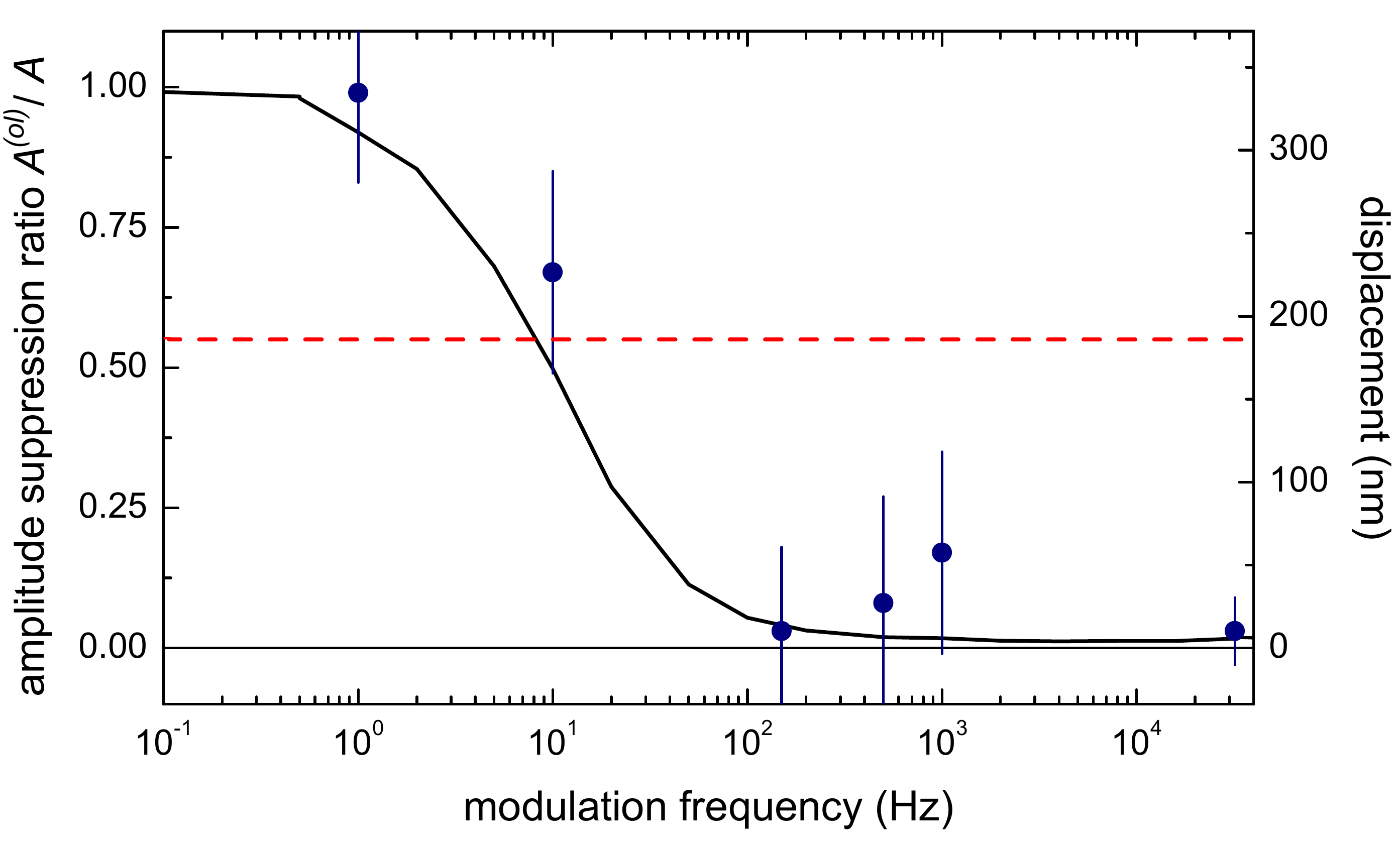}
  \caption{\label{fig:supprvsfreq} Suppression of the ion oscillation amplitude in the lattice, $A^{(ol)}/A$, as a function of the modulation frequency. The dashed line represents a displacement of $\lambda$/2. The results of a numerical simulation for an optical pumping rate of $\Gamma_{op} = 0.1 ~ \Gamma$ and a temperature $T \approx 250 ~\mu$K are shown as a solid line.
  }
  \end{center}
 \end{figure}

In order to compare the suppression of the ion transport by the lattice to a model, we measure the amplitude of the driven ion motion for different drive frequencies. Fig. \ref{fig:supprvsfreq} shows the best repeatedly observed amplitude suppression for each frequency, from which we infer an approximate trapping time on the order of 0.1~s, corresponding to an amplitude suppression of the driven motion by 30$\%$. In order to investigate the ion transport dynamics in the lattice, we use a probabilistic model of the evolution of the ion's energy, which, on a time scale set by the cooling rate, samples by a random walk the thermal distribution of the energies at a finite constant temperature. With a probability per unit time fixed by the cooling rate, the ion energy either increases or decreases by a motional quantum, weighted by the ratio of Boltzmann factors. At any given time, if the ion's energy is smaller than the effective depth of the lattice site it is confined to (modified by the time-dependent external force), the ion stays localized at that site. Otherwise the ion follows the external displacement in the shallow electric potential, until its recapture occurs in one of the optical-lattice sites spanned by the ion’s thermal distribution with accordingly weighted probabilities. As indicated by the solid line in Fig. \ref{fig:supprvsfreq}, we find that both the characteristic trapping time and the approximate frequency dependence of the suppression of ion transport in the experiment are in good agreement with the results of the numerical simulation for a temperature of 250~$\mu$K, corresponding to a mean vibrational quantum number $\aver{n_y}=4$. The increase compared to $\aver{n_{y}}=0.6$ obtained from the Raman spectrum can be explained by the decrease in optical pumping efficiency due to the polarization misalignment that we have introduced to increase the ion scattering rate and the signal-to-noise ratio of the ion position measurement. This increases the recoil heating and is expected to result in a higher temperature, in qualitative agreement with the temperature obtained from the numerical simulation.

In summary, we have demonstrated Raman cooling of an ion in an optical lattice a factor of 60 below the potential depth, a lock-in detection method of the ion position with a resolution of $\lambda / 40$, and trapping of the ion at a single lattice site on millisecond time scales even in the presence of an applied force. These results open new possibilities in various areas: the optical trapping and cooling may enable the realization of a strongly confining hybrid \emph{electrostatic}-optical trap by combining electrostatic forces in two directions with the optical lattice in the third, thereby eliminating the driven micromotion in Paul traps that is disruptive to controlled ion-atom collisions \cite{Grier09,Zipkes10a,Schmid10,Cetina12}, and to scalable quantum computing architectures that rely on ion transport \cite{Blakestad11,Hensinger06}. In combination with the sub-wavelength position-resolution method, this system is also highly suited to experimentally implement simulations of dynamic friction models with ion chains and structural phase transitions of such chains in periodic optical potentials \cite{Pruttivarasin11,Benassi11}.

\begin{acknowledgments}
We thank Yufei Ge, Isaac Chuang and Karl Berggren for assistance with the fabrication of the microchip. This work was supported by the ARO and the NSF. L.K. gratefully acknowledges support by the Alexander von Humboldt-Foundation and A.B. and D.G. gratefully acknowledge support by the NSERC.
\end{acknowledgments}
%

%
\end{document}